\documentclass[11pt,twoside]{article}

\usepackage{galev06}
\usepackage{epsf}
\usepackage{psfig}
\usepackage{lscape}

\begin{document}
\title{Galaxy Evolution through Infrared Surveys: from Spitzer to Herschel}   
\author{A. Franceschini$^1$, M. Vaccari$^1$, S. Berta$^1$, G. Rodighiero$^1$, C. Lonsdale$^2$ }   
\affil{$^1$Astronomy Department, Vicolo Osservatorio 2, I-35122 Padova, Italy} 
\affil{$^2$University of California San Diego, La Jolla, CA 92093-0424, USA} 

\begin{abstract} 
The Spitzer Space Telescope is devoting a significant fraction of the
observing time to multi-wavelength cosmological surveys of different
depths in various low-background sky regions. Several tens of thousand
mid-IR galaxies have been detected over a wide interval of
redshifts. A progressively clearer picture of galaxy evolution is
emerging, which emphasizes populations of luminous galaxies at $z>1$,
likely corresponding to the main phases of stellar formation and
galaxy assembly. These results are entirely consistent with previous
outcomes from ISO, SCUBA and COBE observations, and provide valuable
constraints of high statistical and photometric precision. 
We briefly report here on our attempt to extract from statistical data
some general properties of galaxy evolution and describe evidence that
a population of very luminous objects at $z>1.5$ share different
properties from those of starbursts at lower redshifts, indicating
some seemingly anti-hierarchical behavior of galaxy evolution in the
IR.  We warn, however, that these results are based on large, still
uncertain, extrapolations of the observed mid-IR to bolometric fluxes,
for measuring which the forthcoming far-IR and submillimetre Herschel
Space Observatory will be needed. 
We finally comment, based on our present understanding, about Herschel
capabilities for investigating the early phases of galaxy evolution. 
\end{abstract}


\vspace{-0.5cm}

\section{Introduction}   

The interest of long wavelength observations for cosmological studies
was raised 20 years ago by the  \textit{IRAS} mission, showing the
most luminous local galaxies to emit the bulk of their radiant energy
in the far-IR due to dust reprocessing (e.g. Soifer et al. 1988;
Sanders et al. 1988). Ten years later the \textit{ISO} mission has
found first evidence for strong evolution of dust-enshrouded
starbursts between $z=0$ and $z\sim 1$ (Franceschini et al. 2001;
Elbaz et al. 2002; Genzel \& Cesarsky 2003). Together with parallel
findings in the sub-mm with SCUBA (e.g. Blain et al. 2002) and with
COBE (Hauser et al. 1998), these results have established the
relevance of long wavelength studies for our understanding of galaxy
formation (Franceschini et al. 2003; Baugh et al. 2005): a major
fraction of the emission by the most massive, luminous and short-lived
stars, when they are still embedded inside their parent dusty
molecular clouds, is optically extinguished and reprocessed at IR to
sub-mm wavelengths. 

With the advent of the \textit{Spitzer Space Telescope}, the
exploration of the distant universe at IR wavelengths has become
possible with similar sensitivities and spatial (to some extent also
spectral) resolutions typical of optical searches with large
ground-based telescopes.  \textit{Spitzer} exploits a battery of
sensitive detectors from $\lambda = 3$ to 70 $\mu$m (Rieke et
al. 2004), while its performance at longer wavelengths is limited by
telescope diffraction.  
The 24 $\mu$m band of the MIPS imager, in particular, detects the 8
$\mu$m PAH bundle emission to $z > 2$ for the first time. Thanks to
all this, the observatory has identified large samples of star-forming
galaxies and IR-emitting active galactic nuclei over wide redshift
intervals (Perez-Gonzales et al. 2005; Le Floc'h et al. 2005;
Rowan-Robinson et al. 2005; Caputi et al. 2006; Babbedge et al. 2006;
Dole et al. 2006). 

Early attempts to constrain galaxy evolution based on \textit{Spitzer}
observations have made use of the identifications of large samples of
faint 24 $\mu$m sources with \textit{Spitzer} IRAC near-IR data,
allowing a good characterization of the galaxy SED and photometric
redshifts. In particular, a 0.6 $deg^2$ region of the CDFS was
observed at 24 $\mu$m by Le Floc'h et al. (2005) and a sample of 2600
galaxies brighter than 80 $\mu Jy$ was combined with existing optical
(COMBO17) data in the field and used to derive bolometric IR
luminosity functions and SFR's from z=0 to $\sim1$. These results
imply a comoving IR energy density in the  
Universe to evolve proportionally to $(1+z)^{3.9 \pm 0.4}$ to $z\simeq 1$.
From MIPS 24 $\mu$m observations of the CDFS and HDFN, combined with a
systematic photometric redshift analysis using the Spitzer IRAC data,
Perez-Gonzales et al. (2005) derived estimates of the
redshift-dependent galaxy luminosity functions and found that the SFR
density remains roughly constant above $z=1$.

We will follow a different approach in the next Sects., based on the
analysis of statistical data, whose integral nature is less subject to
the uncertainties in the photometric redshifts estimates.

\begin{figure}
\plottwo{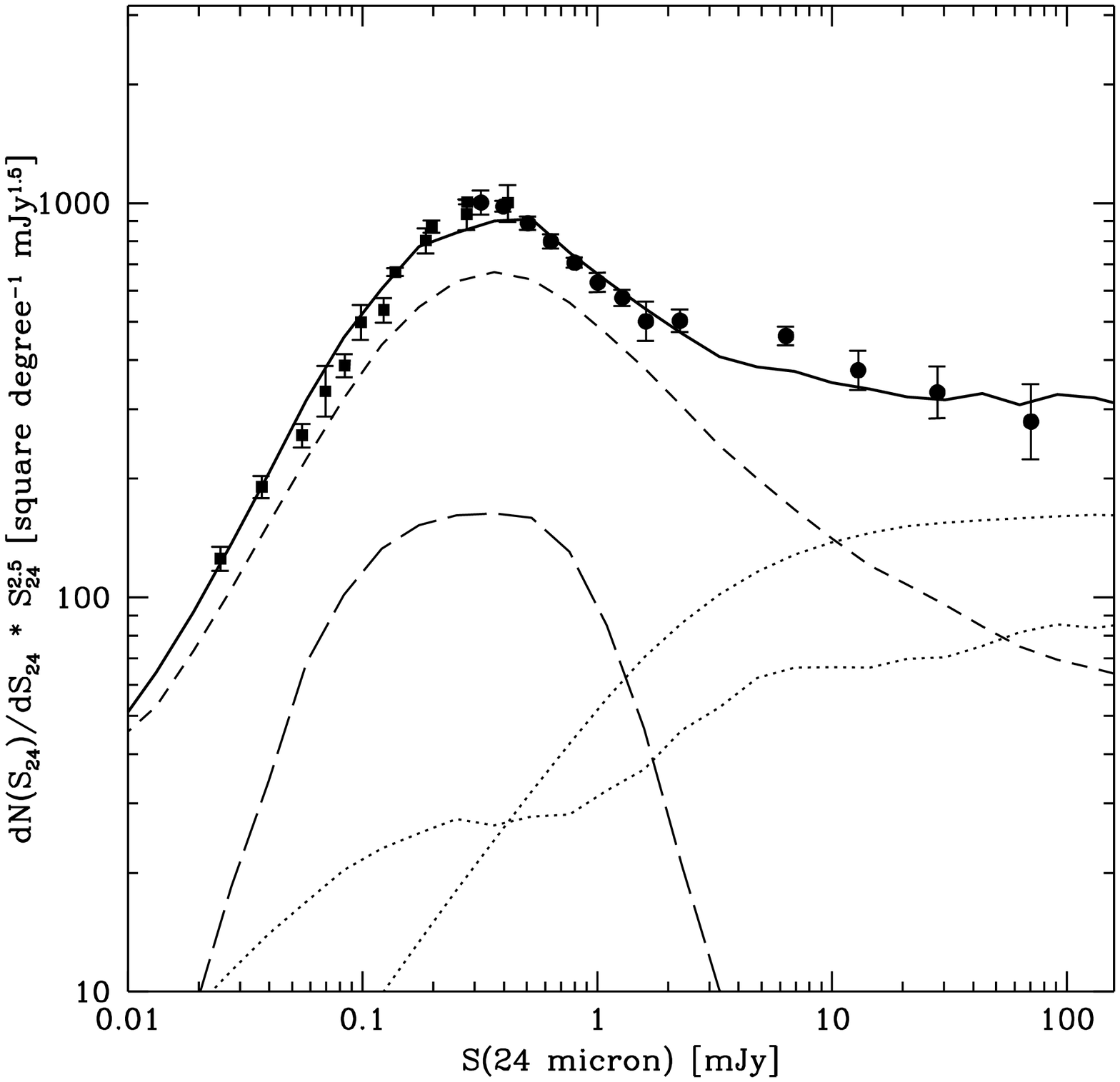}{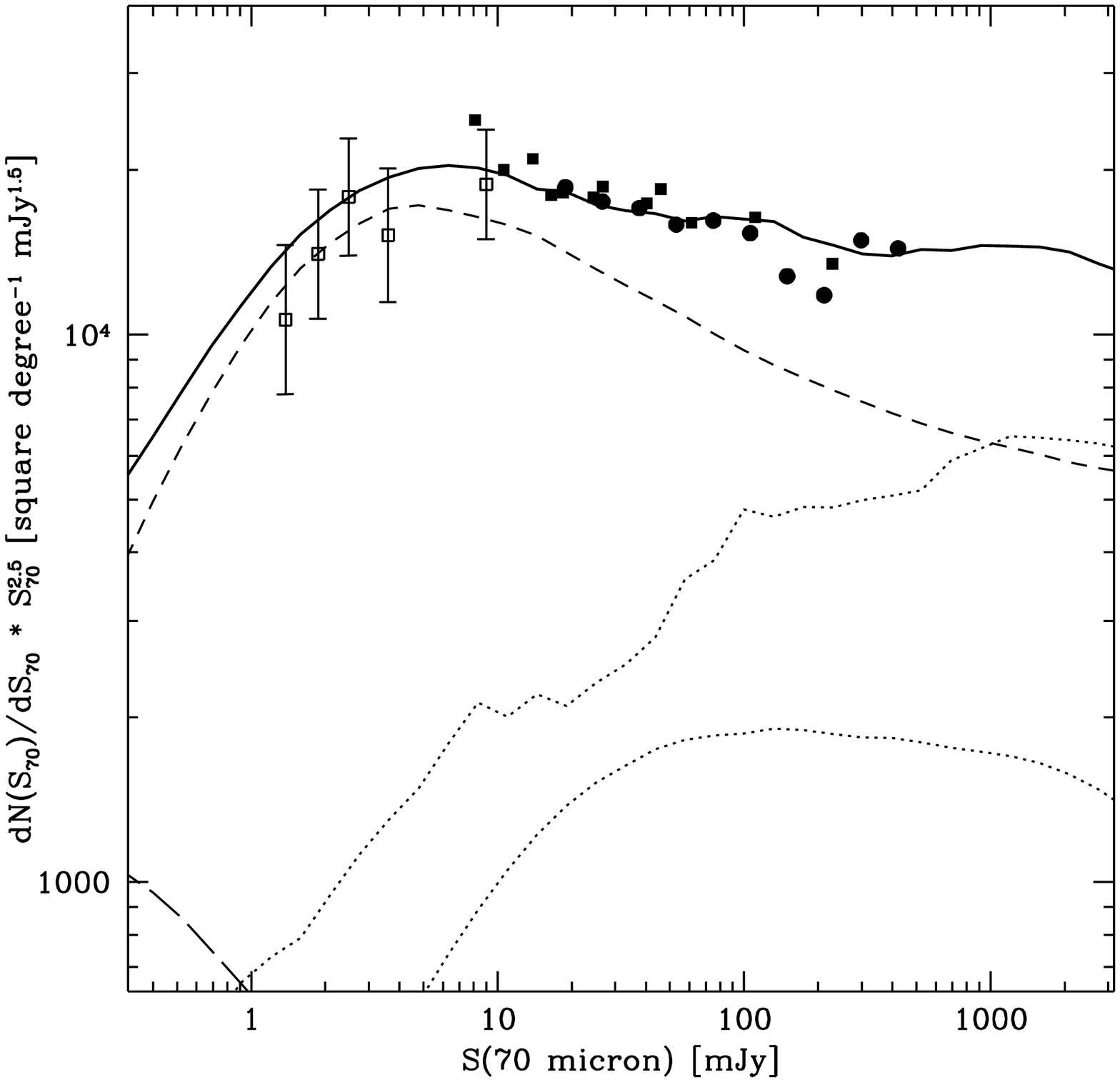}
\caption{Euclidean-normalized differential galaxy counts counts at 24
  and 70 $\mu$m from \textit{Spitzer} MIPS surveys (Sect.2). The red
  dotted line is the model prediction for AGNs, cyan dashes the
  moderate-luminosity and red long-dashes the high-luminosity
  starbursts, black dotted lines normal spirals.} 
\label{c}
\end{figure}

\section{Statistical Observables...}

Galaxy number counts in the most sensitive \textit{Spitzer} MIPS 24
$\mu$m channel have been estimated by various teams with surveys at
different depths. Based on data from the SWIRE survey (Lonsdale et
al. 2004, Rowan-Robinson, these Proceedings), Shupe et al. (2006) have
found quite significant field-to-field variations (up to a factor of
2) at bright fluxes due to local structure.  Averaged over the large
areas covered by SWIRE, the differential counts show a sharply
non-Euclidean behavior as in Figure \ref{c}. Counts at deeper flux
levels (wider cosmic volumes) in smaller fields from Papovich et al
(2004) and Chary et al (2004) reveal a very fast convergence. 
At 70 $\mu$m the bright galaxy counts are from SWIRE (Afonso-Luis et
al. 2006), while the deeper ones from Frayer et al. (2006;
Fig. \ref{c} right panel), and similarly for the 160 $\mu$m data.  

\begin{figure}
\plottwo{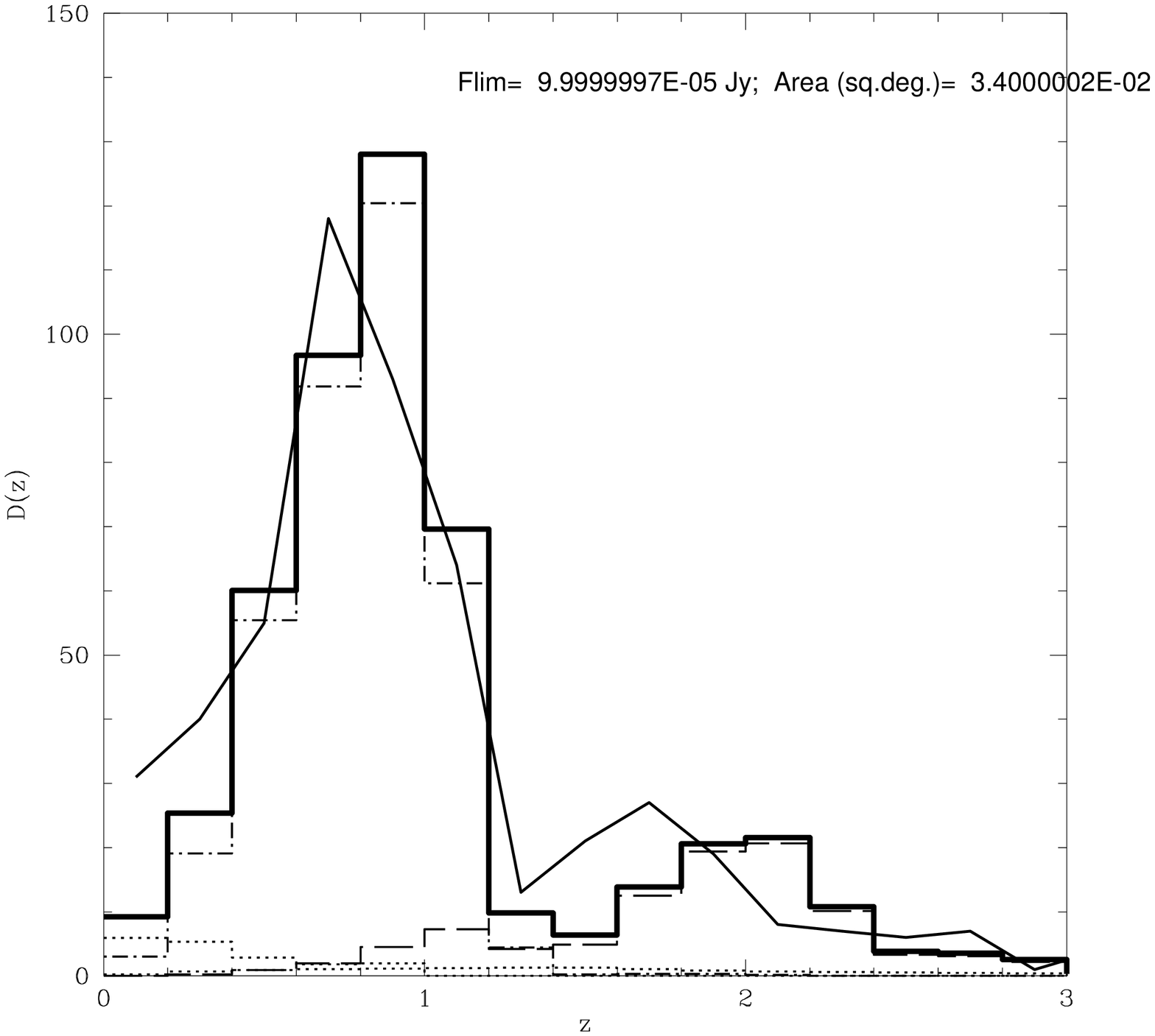}{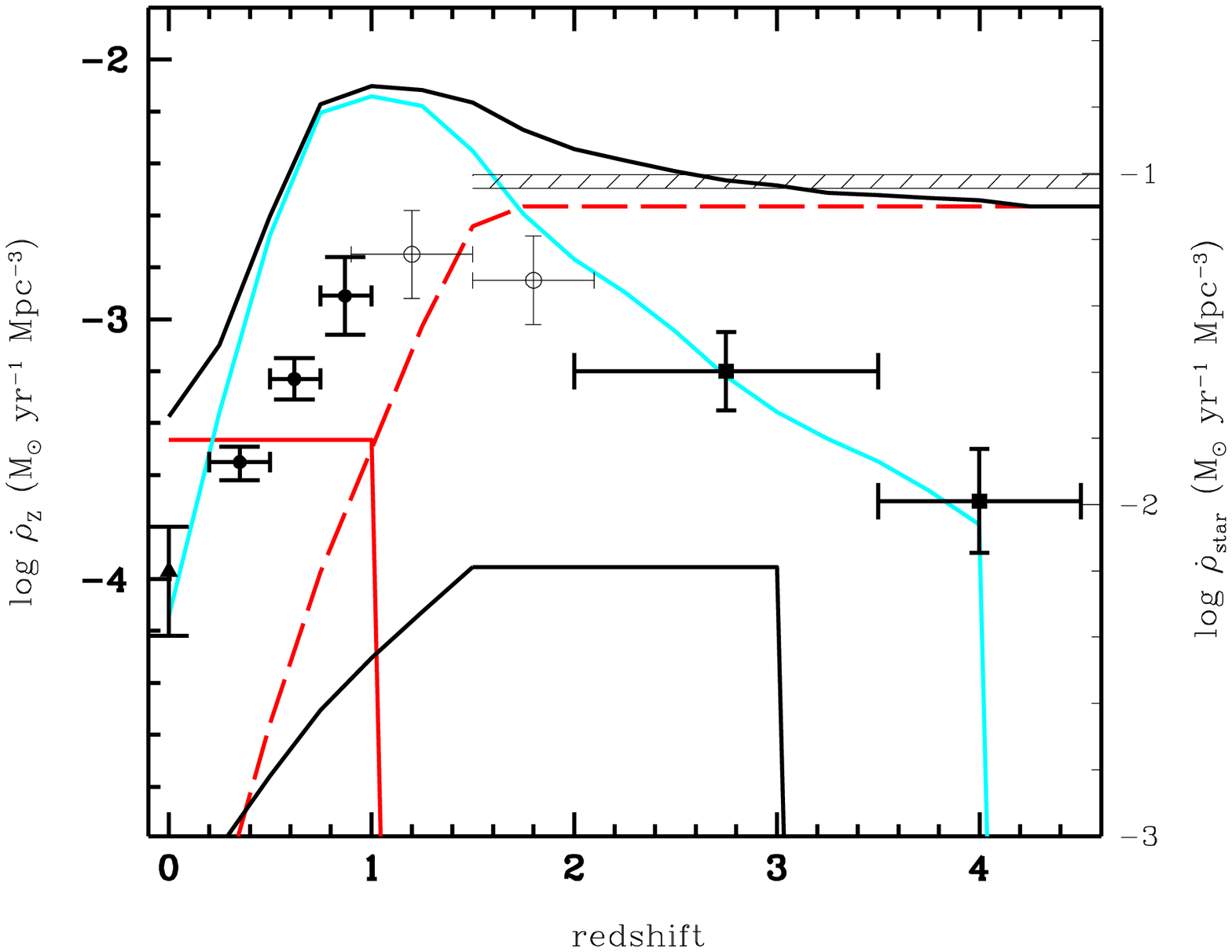}
\caption{\textit{Left panel:} redshift distribution of galaxies in
  GOODS/CDFS brighter than 100 $\mu Jy$ (black broken line) vs. the
  model prediction as discussed in Sect.3 (histograms).  
\textit{Right panel:} evolution of the comoving bolometric emissivity
  (expressed in terms of the equivalent SFR and metal-production rate)
  of various galaxy populations compared with estimates based on
  optically selected galaxy samples. Line types as in Fig.1. 
}
\label{d}
\end{figure}

Another critical constraint is offered by the observed redshift
distributions from complete 24 $\mu$m galaxy samples. Earlier
estimates were reported by Perez-Gonzalez et al. and Caputi et
al. (2005), based on a sample in CDFS flux-limited to 83 $\mu$Jy. We
report in Figure \ref{d} a reassessment of the z-distribution for
galaxies brighter than 100 $\mu$Jy in the CDFS GOODS area (see
http://data.spitzer.caltech.edu/popular/goods), including some new
spectroscopic redshifts and our own estimate of photometric
redshifts. As previously mentioned in Caputi et al., the observed
distribution (broken continuous line) reveals a bimodality, likely due
to the effect of strong PAH emission features in the typical source
SEDs convolved with the MIPS channel transmission function.

Additional relevant data further constraining galaxy evolution come
from the SCUBA number counts and redshift distributions, from the COBE
background intensity and from the IRAS local luminosity functions,
among others.

\section{... and Model Analyses}

This large dataset poses a serious challenge to any attempts to
explain it with simple evolutionary prescriptions. Previous models
(Franceschini et al. 2001; Xu et al. 2004; Pozzi et al. 2004) have
tried to fit the ISO and IRAS data by combining the contributions of
moderately-evolving local spirals with that of a fast evolving
population of dusty starbursts, with evolution rates independent of
luminosity.

\begin{figure}
\psfig{file=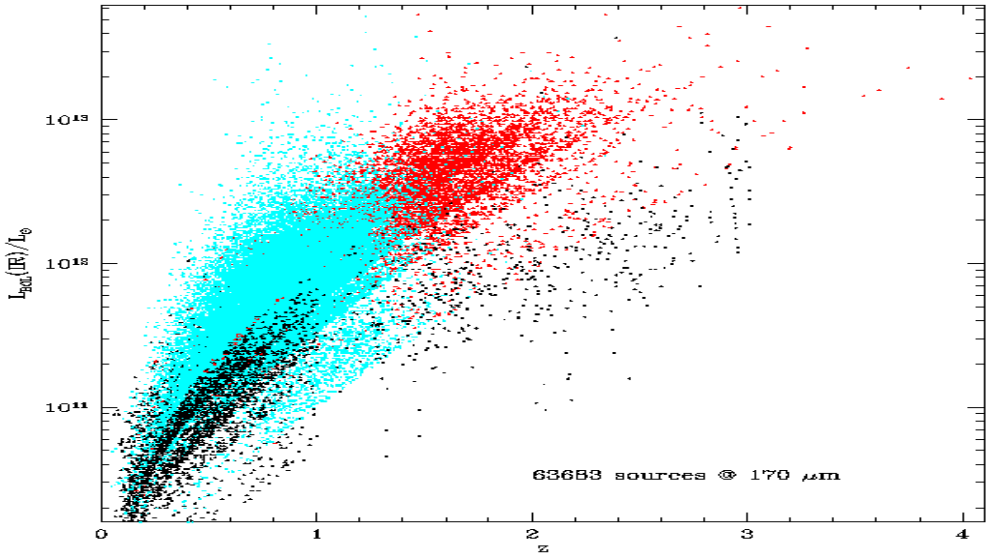,width=14cm,height=8.2cm}
\caption{Simulated outcome of deep galaxy surveys at 170 $\mu$m that
  will be carried out in the Herschel PACS and SPIRE Guaranteed
  Time. The expected bolometric IR luminosity of the detected galaxies
  are plotted as a function of $z$ (colors correspond to galaxy
  populations defined in previous figures).} 
\end{figure}

We have found that this minimal scheme is ruled out by the combination
of the \textit{Spitzer} 24 $\mu$m and SCUBA data. The extremely narrow
peak at fluxes of $\sim$0.3 mJy in the galaxy differential counts
(Fig.\ref{c}) requires a population of moderately luminous starbursts
with maximum comoving IR emissivity around $z=1$ and a fast decline at
higher-z. Such evolutionary rate is illustrated as the cyan line in
the right panel of Fig.\ref{d}, where the bolometric emissivity is
expressed in equivalent SFR density. 
Contributions to the various statistics by this class of sources are
reported as cyan lines in Figs.\ref{c} and \ref{d}.  

The fast decrease of this population at $z>1$ implies that these
objects cannot explain the secondary peak at $z\simeq 2$ in the
redshift distribution of 24 $\mu$m sources and are essentially
unrelated with the high-z sub-millimetric SCUBA population.   
An additional component of very high luminosity starburst galaxies
(ULIRGs), very rare locally but numerous and dominating the emissivity
at $z>1.5$, is then required (red dashed lines in Fig.\ref{c} and
\ref{d}). 

Altogether, the combined set of IR data reveal clear evidence for
\textit{down-sizing} in galaxy evolution: the high luminosity
starburst population (Fig.\ref{d} right) evolved faster at decreasing
redshift and was active at earlier cosmic times, $z>1.5$, compared to
the lower-luminosity objects forming stars for a more prolonged time
and whose main evolutionary phases peak around $z\sim1$. 
This evidence based on IR data (see also Caputi et al. 2006) agrees
with a variety of independent analyses (e.g. Cowie et al. 1996), and
is particularly strong because of the constraint imposed by the
spectral intensity of the far-IR background.

\begin{figure}
\psfig{file=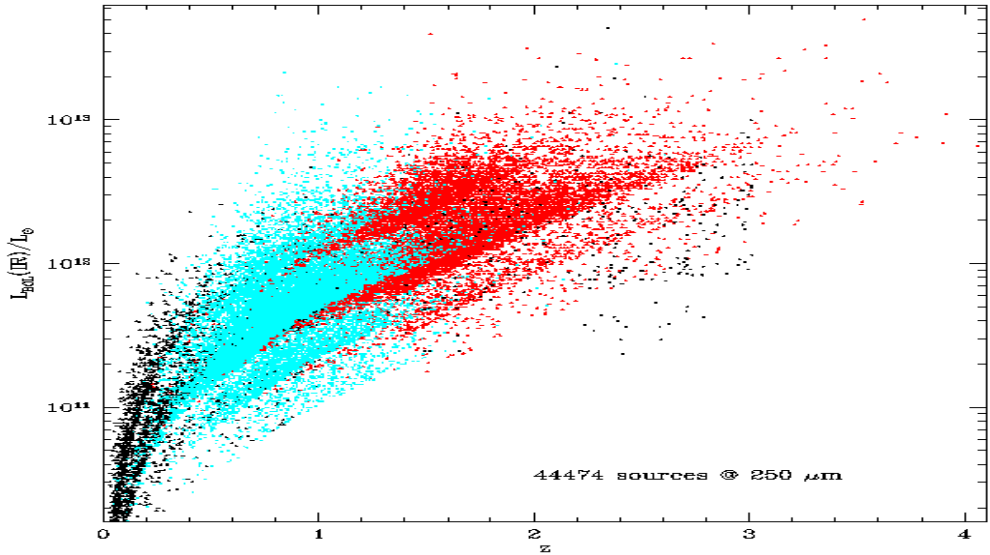,width=14cm,height=8.2cm}
\caption{The simulated IR luminosity vs. $z$ of galaxies that will be
  detected at 250 $\mu$ by the \textit{Herschel} SPIRE instrument in
  the GT observations of various areas and depths. See also the
  previous fig.} 
\end{figure}

\begin{figure}
\psfig{file=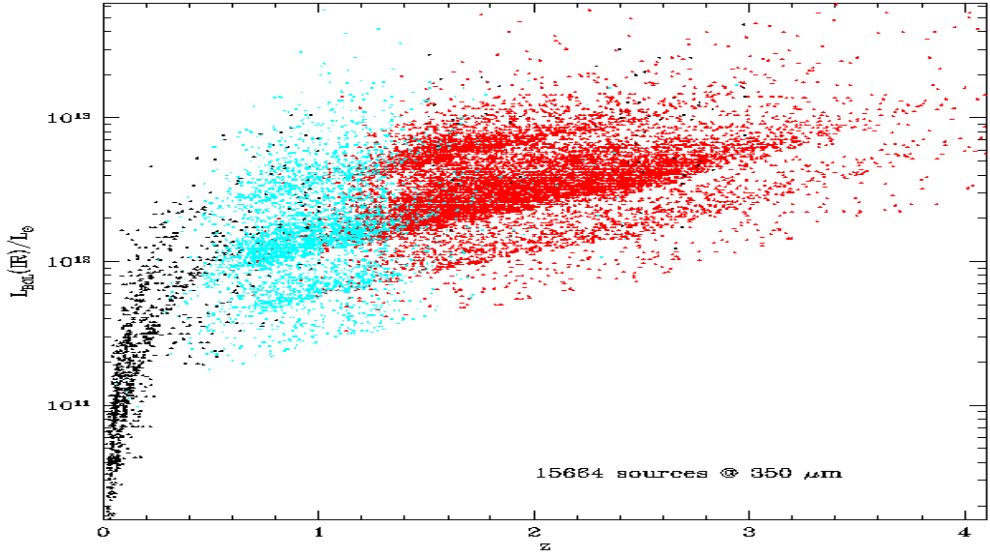,width=14cm,height=8.2cm}
\caption{As in Fig. 4, for galaxies selected by \textit{Herschel} SPIRE at 350 $\mu$.}
\end{figure}

\section{The \textit{Herschel} Perspective}   

Our results on galaxy evolution, as summarized in the right panel of
Fig.\ref{d}, are based on large extrapolations of the observed mid-IR
(and sub-mm) to bolometric fluxes.  In spite of some efforts to
calibrate these relations (e.g. Elbaz et al. 2002), they are still
subject to large uncertainties.  

A major progress is then expected from the forthcoming far-IR and
sub-millimetric \textit{Herschel Space Observatory} mission that,
thanks to the substantial improvement in telescope size and to the
correspondingly lower confusion noise, will allow us to obtain
extensive characterization of the IR to sub-mm SEDs of large samples
of high-z galaxies. The programs for cosmological surveys in the PACS
and SPIRE instrument Guaranteed Times are discussed by Griffin and
Poglitsch (these Proceedings). We exploit here our multi-wavelength IR
model to predict some of the outcomes of such observations. 

Figures 3, 4 and 5 illustrate the expected bolometric luminosity
vs. redshift plots of surveys in various areas to different depths
performed at 170, 250, and 350 $\mu$m. The apparent stripes correspond
to different areas and sensitivity limits of the GT survey program,
and the effects of K-correction are also evident in decreasing the
slope of the L/z correlation at increasing wavelengths. In conclusion,
already within the GT program, the \textit{Herschel} mission will
ensure wide coverage of the galaxy far-IR luminosity functions at
$z<1.5$ and an excellent characterization of the high luminousity end
at higher-$z$.




\end{document}